\documentclass[a4paper,titlepage,twocolumn,superscriptaddress]{revtex4-2}
\usepackage[utf8]{inputenc}
\usepackage{graphicx}
\usepackage{amsmath}
\usepackage{amssymb}
\usepackage{times}
\usepackage[utf8]{inputenc}
\usepackage{nicefrac}
\usepackage{color}
\usepackage{xcolor}
\definecolor{darkblue}{rgb}{0.0, 0.18, 0.65}
\usepackage{hyperref}
\hypersetup{
     colorlinks=true,
     linkcolor=blue,
     filecolor=darkblue,
     citecolor = darkblue,      
     urlcolor=darkblue}

\usepackage{color}
\usepackage{subcaption}
\usepackage{dsfont}
\usepackage{xr} 

\newcommand{\elysee}{\emph{Elys\'ee2017fr}}
\newcommand{\blue}{\textcolor{black}}

\begin{document}

\title{Voter model can accurately predict individual opinions in online populations}
\author{Antoine Vendeville}
\email{antoine.vendeville@sciencespo.fr}
\affiliation{m\'edialab, Sciences Po, 75007 Paris, France}
\affiliation{Complex Systems Institute of Paris Île-de-France (ISC-PIF) CNRS, 75013 Paris, France}
\affiliation{Learning Planet Institute, Learning Transitions unit, CY Cergy Paris University, Paris, France}

\date{\today}

\begin{abstract}
     Models of opinion dynamics describe how opinions are shaped in various environments. While these models are able to replicate \blue{general} opinion distributions observed in real-world scenarios, their capacity to align with data at the \blue{user} level remains mostly untested. We evaluate the capacity of the multi-state voter model with zealots to capture individual opinions in a fine-grained Twitter dataset collected during the 2017 French Presidential elections. Our findings reveal a strong correspondence between individual opinion distributions in the equilibrium state of the model and ground-truth political leanings of the users. Additionally, we demonstrate that discord probabilities accurately identify pairs of like-minded users. These results emphasize the validity of the voter model in complex settings, and advocate for further empirical evaluations of opinion dynamics models at the \blue{user} level.
\end{abstract} 

\maketitle

\section{Introduction} 

Models of opinion dynamics provide potent tools to unveil mechanisms driving phenomena such as polarization \cite{peralta2024multidimensional,ojer2023modeling,pham2022empirical,baumann2021emergence,peralta2021effect,axelrod2021preventing,macy2021polarization} or echo chambers \cite{vendeville2024echo,vendeville2023opening}. These models usually comprise a network of agents endowed with opinions that evolve through interactions with one another. Social, psychological and technological features may govern these interactions---e.g.\ negative influence \cite{pham2022empirical,axelrod2021preventing,macy2021polarization,li2015voter}, stubbornness \cite{vendeville2023discord,yildiz2013binary,mobilia2007role} or recommendation algorithms \cite{iannelli2022filter,ramaciottimorales2021auditing,peralta2021effect}. Exploring the effect of such features and combinations thereof, can help us understand how ideological landscapes are shaped in various circumstances. This is particularly useful in the current context of raising concerns about the role of online social platforms in political debates and democratic processes \cite{vishnuprasad2024tracking,lorenzpreen2023systematic,kubin2021role}.

The capacity of the models to replicate real-world observations has garnered increasing attention in recent years. Recurrent sources of data employed towards this goal include surveys \cite{ojer2023modeling,perezmartinez2023polarized,baumann2020modeling,sikder2020minimalistic}, election results \cite{gsanger2024opinion,meyer2024time,vendeville2021forecasting,mori2019voter,gracia2014is}, and interactions extracted from online social platforms \cite{peralta2024multidimensional,liu2023emergence,valensise2023drivers,baumann2021emergence,sasahara2020social}. The celebrated voter model has been the subject of several studies pertaining to this question. The seminal paper of Fern\'andez-Gracia et al.\ \cite{gracia2014is} demonstrated the model's ability to capture statistical features of vote shares in U.S.\ elections. These results were confirmed and enriched in subsequent works \cite{braha2017voting,mori2019voter,vendeville2021forecasting,gsanger2024opinion,meyer2024time}. We refer the interested reader to \cite{peralta2022opinion,sobkowicz2020whither} for in-depth reviews of the research about the empirical validity of opinion dynamics models.

\blue{Models are opinion dynamics are largely used to analyze global phenomena, such as polarization or radicalization. Therefore, research about the empirical validity of the models usually stands from a global point of view. For example, Valensise et al.~\cite{valensise2023drivers} investigate the capacity of four different opinion dynamics models to replicate global opinion distributions observed in various online social platforms, while the model employed by Peralta et al.~\cite{peralta2024multidimensional} is able to reproduce multidimensional opinion distributions of X/Twitter (hereafter Twitter) users. The capacity of the models to identify opinions at the individual level remains mostly untested. It is particularly crucial to evaluate this capacity, because of the horizontal and heterogeneous nature of online social platforms. The political debate in traditional media was structured in a vertical manner, with small number of pundits expressing their opinions. But online platforms have placed the individual at the center of the debate, with the intent that any voice should be able to express itself and be heard by others. In parallel, a growing emphasis has been put on individual differences and particularities in Western societies \cite{rosanvallon2021populist}. Therefore, the heterogeneity of users is a crucial component of online opinion dynamics, which warrants specific attention and models that take it into account.}

\blue{The lack of empirical evaluations of opinion dynamics models at the individual level owes to a two-fold difficulty. The first is the inherent cost associated with the collection of fine-grained data suitable for this type of analysis, even moreso now that the platforms traditionally considered for these types of studies---Twitter and Reddit---have locked access to their data behind expensive paywalls \cite{roozenbeek2022democratize}. The second and perhaps more challenging difficulty is the estimation of model parameters from data. It is not straightforward to operationalize and measure features such as confirmation bias, stubbornness, or the influence of algorithmic recommendations. While such parameters can sometimes be estimated on the basis of prior information \cite{sasahara2020social}, empirical applications of opinion dynamics models often rely on sweeps of the parameter space to find values for which the model best reproduces the observed data \cite{peralta2024multidimensional, valensise2023drivers, pansanella2023mass, sasahara2020social}. For the sake of robustness, these analyses should ideally be complemented with evaluations of the models thus calibrated on different, but comparable, datasets. This is no easy task, given the diversity in the nature of the datasets, and the inherent cost associated with obtaining multiple of them.} 

\blue{We address these issues as we evaluate the capacity of the voter model to capture individual opinions in a fine-grained real-life dataset. Our methodology is parameter-free and thus circumvents the difficulty of parameter inference. We leverage a directed, weighted network of retweets between Twitter accounts, collected during the campaign of the 2017 French Presidential Election \cite{elysee2017_data}. Treating accounts of political entities as static reference points (zealots), recent theoretical advances on the multi-state voter model in complex networks \cite{vendeville2023discord} allow us to infer individual opinion distributions $x_i$ at equilibrium. We then compare these theoretical opinion distributions with ground-truth party affiliations $y_i$ explicitly stated by the users in their publications and self-descriptions. We uncover a strong correspondence between the two. Supporters of the same party are clustered together in the multi-dimensional space spanned by the $x_i$ vectors, clearly separated from those with different views. The mode of $x_i$, which indicates the opinion most often held by user $i$ in any realization of the model, matches the ground-truth $y_i$ in 92.5\% of cases. We also show that discord probabilities---which quantify the frequency at which any two users disagree in the voter model---let us deduce with high accuracy whether or not two users support the same party. Neither the follow nor the mention networks yield comparable results, while the undirected or unweighted counterparts of the retweet network approach this performance but do not equate it. Details of the computations, networks statistics, and additional figures can be found in the Supplemental Material.}

\section{Methodology} 

The multi-state voter model with zealots for complex networks unfolds as follows \cite{vendeville2023discord}. Consider a directed, weighted network of users labelled $1,\ldots,N$ with weight $w_{ij}$ on the edge $j\rightarrow i$. Users are initially endowed with a random opinion chosen in a finite discrete set $\mathcal{S}$. In iterated steps, a user $i$ is chosen at random and adopts the opinion of another user chosen at random according to the probabilities $\{w_{ij}\}_{j=1,\ldots,N}$ (edge weights are normalized so that each user has in-degree 1). Let us suppose that users who never change opinion, called zealots \cite{meyer2024time,vendeville2023discord,ramirez2022local,mobilia2007role}, are present in the system. Let $\mathcal{N}$ denote the set of non-zealot users, and $z_i^s$ the aggregated weight of edges from zealots with opinion $s$ toward user $i\in\mathcal{N}$. As long as every user $i$ can be reached by a zealot (either $z_i^s>0$, or there exists an ancestor $j$ of $i$ such that $z_j^s>0$), the system converges towards a unique state of equilibrium \cite{yildiz2013binary}. In this state, the opinion of $i$ fluctuates according to an individual probability distribution $x_i$ such that, for any $s\in\mathcal{S}$ \cite{vendeville2023discord},
\begin{equation} \label{eq:xis}
     x_i^s = \sum_{j\in\mathcal{N}} w_{ij}x_j^s + z_i^s.
\end{equation}
\blue{Importantly, this value represents an average at equilibrium, and does not inform on the precise dynamics of a single realization of the model.}

Our methodology consists in applying this model on the \emph{\#Elysée2017fr} dataset \cite{elysee2017_data}. The creators of the dataset collected retweets and mentions between 22,853 Twitter accounts in the last six months leading to the 2017 French Presidential Election. Follow relationships between the accounts were collected in another study \cite{papanastasiou2023}. The accounts were selected based on the presence of political keywords in their tweets, and include about 2,000 non-individual political entities, e.g.~official party accounts, activist groups, etc. On the basis of the content of their tweets and profile description, accounts were manually labelled by the creators of the dataset to indicate the party they support in the election: FI (France Insoumise, far-left), PS (Parti Socialiste, traditional left), EM (En Marche, center), LR (Les Républicains, traditional right), FN (Front National, far-right). These were the five main competing parties in the elections. We discard unlabelled accounts and accounts with multiple labels (resp.\ 3,401 and 803 accounts). \blue{The reliability of these labels was demonstrated by the creators of the dataset in the corresponding article \cite{elysee2017_paper}, with an observed agreement of 89\% between annotators.}

To apply the voter model in this context, we consider an opinion space formed of the different parties: 
\begin{equation}
     \mathcal{S}=\{FI,PS,EM,LR,FN\}.
\end{equation}
Each account is endowed with a ground-truth opinion $y_i\in\mathcal{S}$, indicating the party they support as identified by the creators of the dataset. We call supporters of party $s$ all users with ground-truth opinion $y_i=s$. Retweet interactions induce a directed, weighted network so that the weight $w_{ij}$ of the edge from $j$ to $i$ is defined as \blue{the proportion of $i$'s retweets for which the original tweet is from $j$}. We treat accounts of political entities as zealots and fix their opinions to their ground-truth opinions, which correspond to the party they officially represent. The value $z_i^s$ is thus proportional to the number of times that $i$ has retweeted a political entity that represents party $s$. 

A central aspect of our methodology is to consider political entities as zealots. It seems natural to assume that party affiliations of political entities are immutable: the Parisian branch of FI for example, is and will always be---by definition---in support of FI. As these affiliations are public knowledge, it is not such a restrictive hypothesis to assume that they are known. This choice is also motivated by a technical reason: zealots guarantee the existence of a unique equilibrium state, at least for the regions of the network that they can reach \cite[Thm.~2.1]{yildiz2013binary}. Otherwise, there may exist a multitude of different equilibrium states, the potential diversity of which entails a difficulty in their interpretation. Zealots also act as reference points, ensuring a unique correspondence between  opinions in the model and ground-truth opinions. 

\blue{We evaluate the capacity of the model thus defined to retrieve ground-truth opinions. Those of the zealots are already used as reference points, so the performance of the model is evaluated on non-zealot users, which we call users for the sake of simplicity.} We compute individual equilibrium opinion distributions as per Eq.~\ref{eq:xis}. In other words, the value of $x_i^s$ can be interpreted as the likeliness attributed by the model to the possibility that $i$ supports party $s$. Crucially, the ground-truth opinion $y_i$ of $i$ does not bear involvement in the computation of $x_i^s$. This justifies our analysis, where we compare $x_i$ and $y_i$ in several fashions to assess the validity of the model.

\section{Results}

\subsection{Equilibrium opinion distributions}
\begin{figure}[!t]
     \centering
     \includegraphics[width=.48\textwidth]{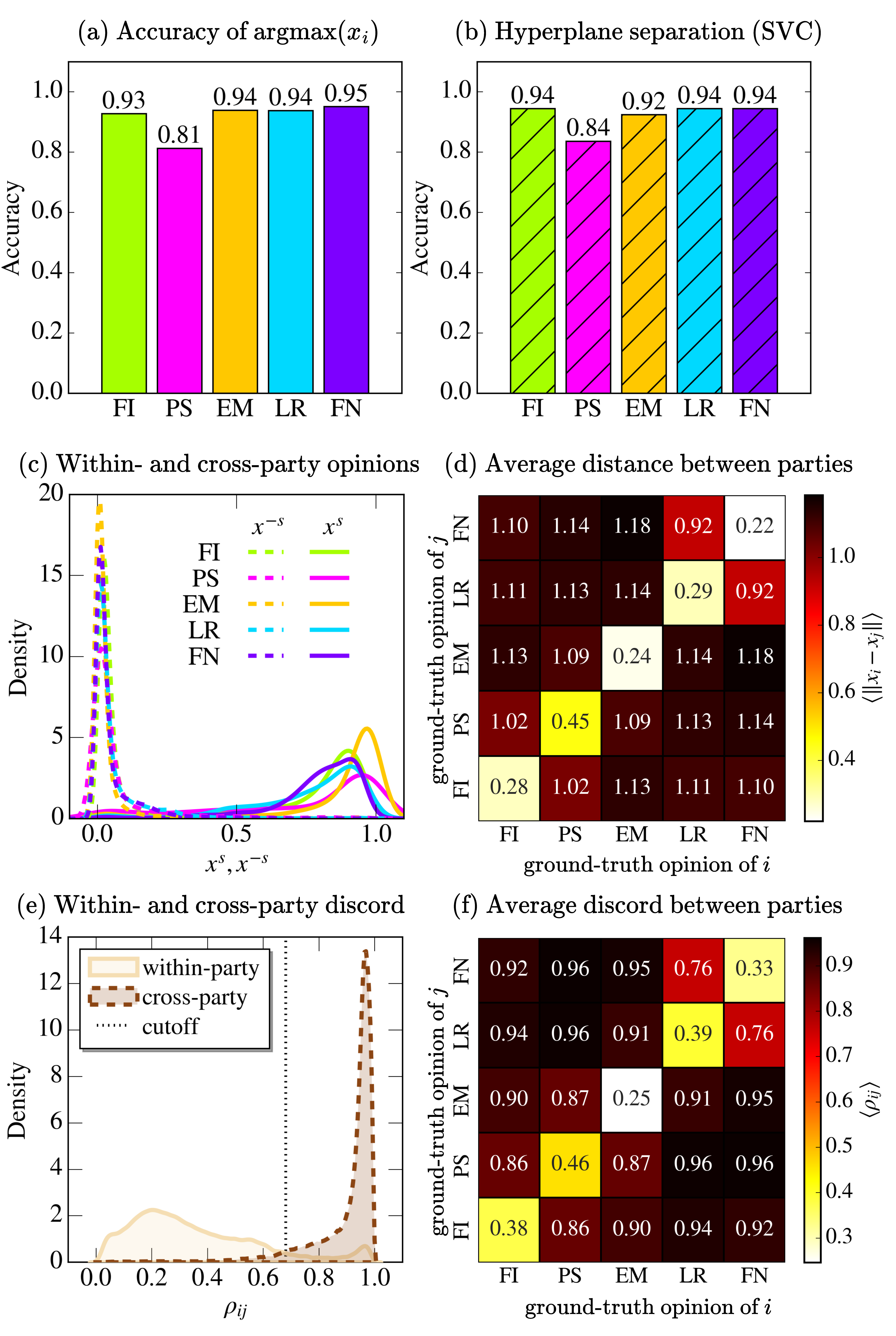}
     \caption{Correspondence between theoretical opinion distributions $x_i$ and ground-truth opinions $y_i$. \textbf{(a)} Party-wise accuracy of argmax$(x_i)$ (plain bars) \blue{\textbf{(b)} Accuracy of the hyperplane separation computed by the SVC algorithm.} \textbf{(c)} Distribution of within-party $(x^s)$ and cross-party $(x^{-s})$ opinions among supporters of each party $s$, distinguished by color. \textbf{(d)} Average euclidean distance between supporters of the same party (diagonal cells) and supporters of different parties (off-diagonal cells). \textbf{(e)} Distribution of discord probabilities $\rho_{ij}$ between supporters of the same party (blue) and supporters of different parties (red, dashed). The vertical dotted line indicates the cutoff of the logistic regression model: if $\rho_{ij}$ is on the left of it, the logistic regression predicts that $i$ and $j$ support the same party, otherwise it predicts that they support different parties. \textbf{(f)} Average discord probability between supporters of the same party (diagonal cells) and supporters of different parties (off-diagonal cells). Smoothed distributions are computed via kernel density estimation.}
     \label{assemble}
\end{figure}

The model is able to identify supporters of each party with very high accuracy. The most likely opinion of $i$ according to the model, i.e.\ $\text{argmax}(x_i)$, matches the ground-truth $y_i$ for 92.5\% of users. Party-wise accuracy values are shown in Fig.~\ref{assemble}a. FN supporters are the easiest to identify (accuracy 0.95), while PS supporters are the hardest (accuracy 0.81). The worse performance for PS is a pattern that will repeat throughout our analysis. It might be due to the dire situation of the PS at the time, which was considerably weakened after five years of difficult Hollande presidency, and its voting base torn between the rise of FI on the left and EM on the right. 

Not only does the model attributes a higher probability to ground-truth opinions than to others, but the difference between the probabilities is not negligible. Let $\mathcal{N}_s$ denote the set of supporters of party $s$. In Fig.~\ref{assemble}c we compare the distributions of
\begin{align}
     x^s &= \{x_i^s: i\in\mathcal{N}_s\}, \text{ and}\\ 
     x^{-s} &= \{x_i^t: i\in\mathcal{N}_s, t\neq s\}. 
\end{align}
The two are concentrated as opposite sides of the unit interval. The former is located towards 1, with means ranging from $\langle x^s\rangle=0.71$ for PS to $\langle x^s\rangle=0.86$ for EM, while the latter is strongly skewed towards 0, with means ranging from $\langle x^{-s}\rangle=0.03$ for EM to $\langle x^{-s}\rangle=0.07$ for PS. Note that we find consistently low probabilities for $x^{-s}$ (standard deviation ranging from 0.09 for FN supporters to 0.15 for PS), while the magnitude of $x^s$ is much more varied (standard deviation between 0.16 for FN supporters and 0.31 for PS).
\begin{figure}[!t]
     \centering
     \includegraphics[width=.48\textwidth]{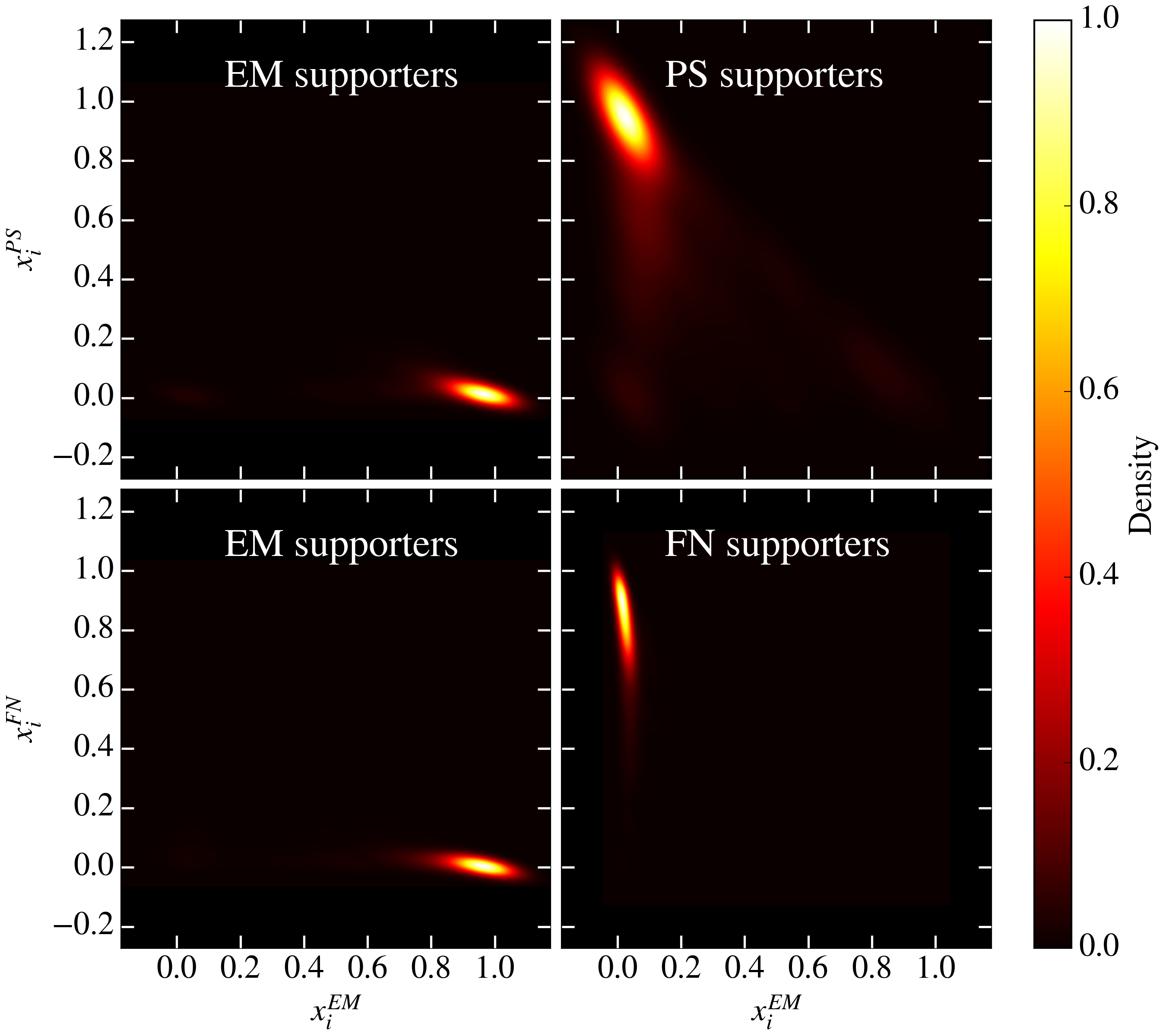}
     \caption{\textbf{(Top)} Distributions of theoretical opinions towards EM and PS, among EM supporters (left) and PS supporters (right). \textbf{(Bottom)} Distributions of theoretical opinions towards EM and FN, among EM supporters (left) and FN supporters (right).}
     \label{2d_projection}
\end{figure}

An interpretation of the striking difference between the distributions of $x^s$ and $x^{-s}$, is that the supporters of each party occupy a specific region in the five-dimensional space spanned by the $x_i$ vectors. Figures~\ref{assemble}d shows that in this space, users are significantly closer to those who support the same party than to others. Furthest away are the supporters of EM and FN, and closest to one another are those of LR and FN. As an illustrative example, we show in Fig.~\ref{2d_projection} the bidimensional distributions of $(x_i^\text{EM},x_i^\text{PS})$ for EM and PS supporters, and $(x_i^\text{EM},x_i^\text{FN})$ for EM and FN supporters. In the latter case, the densities are concentrated at opposite sides of the unit square. The former case however presents a different picture, as supporters of PS populate a larger and more central area of the subspace (Fig.~\ref{2d_projection}, top plots). \blue{To confirm these findings, we evaluate in a more systematic manner how easily supporters of different parties can be separated in the space spanned by the $x_i$ vectors. By fitting a basic Support Vector Classifier (SVC) to the $(x_i,y_i)$ pairs, we find the best separating hyperplane between supporters of the different parties. The separation induced by the hyperplane groups 93\% of the users with their like-minded peers; we also show party-wise accuracies in Fig.~\ref{assemble}b.}

In summary, through the individual equilibrium distributions $x_i$, the voter model embedds the users in a space where supporters of each party are located in a specific region, separated from the others, and therefore clearly identifiable. \blue{This is particularly remarkable, given that individual party preferences are not taken into account by the model.} 

\subsection{Discord probabilities} 

Individual equilibrium opinion distributions $x_i$ describe the stabilization of the system at the user level. It is often useful to also describe stabilization at the level of edges, or more generally user pairs, via the notion of active links or discord probabilities \cite{vendeville2023discord,ramirez2022local,vazquez2008analytical}. The discord probability $\rho_{ij}$ is the probability to find $i$ and $j$ holding two different opinions. Under the same conditions as for $x_i$, this quantity converges towards an equilibrium state
\begin{align}
     \rho_{ij} = \frac{1}{2} &\left[ \sum_{k\in\mathcal{N}} w_{ik}\rho_{jk} + \sum_{s\in\mathcal{S}} z_i^s(1-x_j^s) \right.\notag\\ 
     &\left.+ \sum_{k\in\mathcal{N}} w_{jk}\rho_{ik} + \sum_{s\in\mathcal{S}} z_j^s(1-x_i^s)\right]\label{eq:rho}.
\end{align}
\blue{The first two sums quantify the rates at which user $i$ adopts an opinion different than $j$'s via, respectively, the influence of other users and the influence of zealots. Similarly, the last two sums quantify the rate at which user $j$ adopts an opinion different than $i$'s. Certain conditions of independence between the opinions of $i$ and $j$ allow to use simply 
\begin{equation} \label{rhoij_indep}
     \rho_{ij} = \sum_{s\in\mathcal{S}} x_i^s(1-x_j^s).
\end{equation}
Namely, if either (i) $\sigma_i$ or $\sigma_j$ is constant, or (ii) there is no path from $i$ to $j$ nor from $j$ to $i$, and $i$ and $j$ have no common ancestor. The derivation of Eqs.~\ref{eq:rho},\ref{rhoij_indep} is detailed in ref.~\cite{vendeville2023discord}. As long as the existence and unicity of the $x_i$ vectors are guaranteed, the same goes for the $\rho_{ij}$ values. }

The value $\rho_{ij}$ can be interpreted as the likeliness attributed by the model to the possibility that $i$ and $j$ do not support the same party. Averages $\langle\rho_{ij}\rangle$ within and between parties are shown in Fig.~\ref{assemble}f. EM supporters exhibit the lowest within-party discord, while PS exhibits the highest. The lowest cross-party discord is observed between LR and FN, the highest between PS and both LR, FN. In Fig.~\ref{assemble}e, we plot separately the within-party and cross-party distributions of $\rho_{ij}$. The two differ greatly, the latter being located towards 1 and the former towards 0. Users with the same ground-truth opinion are much more likely to hold the same opinion in the model than those who do not. In fact, a simple logistic regression model is able to discriminate between within-party and cross-party pairs, based on the values $\rho_{ij}$, with 93\% accuracy. This highlights the capacity of the model to reliably distinguish friends from foes. Because discord spans a narrower range cross-party than within-party, foes are easier to identify than friends (accuracy 93.4\% versus 90.2\%).

\section{Robustness}

\subsection{Estimating opinions directly from zealots}

The value of $x_i^s$ corresponds to the probability that a backward random walk initiated in $i$ hits a zealot with opinion $s$ before a zealot with any other opinion \cite[Prop.~3.2]{yildiz2013binary}. Therefore zealots---here political entities---play an crucial role in the equilibrium state of the model. One may then wonder, are the encouraging results we just exposed a mere consequence of the fact that people are preferentially connected with political entities of the party they support? The question is particularly relevant, given that retweet networks are known to exhibit political homophily \cite{garimella2017longterm,halberstam2016homophily,conover2011political}. Let $z_i=(z_i^s)_{s\in\mathcal{S}}$ denote the vector containing the weight of all direct connections from zealots with opinion $s$ toward user $i$. This vector describes the tendency of $i$ to retweet political entities affiliated to the different parties. We expect the vectors $z_i$ to provide good indicators of ground-truth opinions; but if they were to be as good or even better than $x_i$, one may question the relevance of using Eq.~\ref{eq:xis}. 

Because 27\% of users do not have direct connections with zealots, one cannot infer anything from $z_i$ for about a quarter of the considered population. Therefore the use of the vectors $x_i$ is already beneficial in that it extends the realm of the analysis. Let us now focus on the 73\% of users who exhibit direct connections with zealots. Via the $z_i$ vectors, supporters of the same party are positioned closer to one another (average Euclidean distance $\langle\Vert x_i-x_j\Vert\rangle=0.190$) and further from supporters of other parties ($\langle\Vert x_i-x_j\Vert\rangle=1.339$) than they are via the $x_i$ vectors ($\langle\Vert x_i-x_j\Vert\rangle=0.219, 1.146$ respectively). However, we find that $x_i$ is more reliable than $z_i$ to identify ground-truth opinions. Argmax$(x_i)$ matches the ground-truth opinion of $i$ for 96.2\% of users (95.1\% for argmax$(z_i)$) and a Support Vector Classifier fitted on the $(x_i,y_i)$ pairs correctly categorizes 96.4\% percent of the users (94.8\% when fitted on the $(z_i,y_i)$ pairs). Therefore, while direct connections with zealots separate more clearly between supporters of different parties, the voter model allows for a more accurate identification of individual opinions.

\subsection{Comparison with other networks}

Multiple types of user networks can be extracted from online social interactions. To put our results into perspective, we investigate whether different data collection and processing methods impact the empirical validity of the model. We consider five other networks: the unweighted (Unw.)\ counterpart of the retweet network, the undirected (Und.)\ counterpart of the retweet network, the undirected and unweighted counterpart of the retweet network (UU), the Follow network (where $w_{ij}=1$ if $i$ follows $j$ and 0 otherwise), and finally the Mention network (where $w_{ij}$ is the number of times that $i$ mentioned $j$ in a tweet). \blue{The distribution of ground-truth opinions among users and zealots varies very little between the different networks (Fig.\ 1 of the Supplemental Material).}

In Table~\ref{tab:allgraphs} we show that the retweet network performs the best according to all the metrics we have used in the above analysis, except for the within-party distance between users $\langle\Vert x_i-x_j\Vert\rangle_\text{within}$. The poorer performances of the follow and mention networks are not too surprising. Indeed, while retweets are often considered to be markers of endorsement, people may follow a broader range of the political spectrum \cite{garimella2017longterm,halberstam2016homophily}, and mentions are often used to express hostility towards outgroup members \cite{tacchi2022signed,williams2015}. In fact, only 2\% of FN supporters are correctly identified by argmax$(x_i)$ in the mention network. The combination of discarding both weights and directionality (UU network) also affects the performance significantly---only 6\% of FN supporters are correctly identified by argmax$(x_i)$ in the UU network. This is an especially relevant finding, given that a large part of the literature on the voter model focuses on undirected unweighted networks. \blue{The performance of the UU, Follow and Mention networks may also be affected by the fact that zealots bear less overall influence in these networks than in the others (Figs.\ 2,3 of the Supplemental Material). }

\blue{Finally, the unweighted and undirected counterparts of the retweet network exhibit only slightly worse performance than the original retweet network with weights and directionality of edges. Moreover, discarding edge weights appears to be less penalizing that discarding directionality.}

\begin{table}[t]
     \begin{tabular}{lcccccc}
          \hline
          & Retweet & Unw.\ & Und.\ & UU & Follow & Mention \\ \hline
          Acc.\ argmax$(x_i)$ & \textbf{92.5} & 91.6 & 91.9 & 74.9 & 84.2 & 64.4 \\
          Acc.\ SVC & \textbf{93.0} & 92.8 & 92.8 & 92.8 & 89.4 & 83.9 \\
          $\langle x^s\rangle$ & \textbf{0.781} & 0.672 & 0.636 & 0.507 & 0.471 & 0.337 \\
          $\langle x^{-s}\rangle$ & \textbf{0.055} & 0.082 & 0.091 & 0.123 & 0.132 & 0.166 \\ 
          $\langle\Vert x_i-x_j\Vert\rangle_\text{within}$ & 0.296 & 0.266 & 0.235 & 0.188 & 0.205 & \textbf{0.164} \\
          $\langle\Vert x_i-x_j\Vert\rangle_\text{cross}$ & \textbf{1.097} & 0.904 & 0.840 & 0.617 & 0.545 & 0.301 \\
          $\langle\rho_{ij}\rangle_\text{within}$ & \textbf{0.355} & 0.486 & 0.499 & 0.591 & 0.638 & 0.682 \\
          $\langle\rho_{ij}\rangle_\text{cross}$ & \textbf{0.902} & 0.852 & 0.835 & 0.779 & 0.795 & 0.730 \\
          Acc.\ logistic$(\rho_{ij})$ & \textbf{92.9} & 89.8 & 87.2 & 86.1 & 85.3 & 77.8 \\
         \hline
     \end{tabular}
     \caption{Comparison between the results of the model applied on the directed weighted retweet network, its unweighted counterpart (Unw.), its undirected counterpart (Und.), its undirected unweighted counterpart (UU), the follow network and the mention network. Subscripts precise whether the averages are within- or cross-party. $\langle\Vert x_i-x_j\Vert\rangle$ is the average Euclidean distance in the opinion space of the model for the considered user pairs, and $\langle\rho_{ij}\rangle$ the average discord probability. logistic$(\rho_{ij})$ denotes a logistic regression model trained to distinguish between supporters of the same party and supporters of different parties on the basis of the $\rho_{ij}$ values. ``Acc.'' stands for accuracy, given in percentages.}
     \label{tab:allgraphs}
\end{table}

\section{Discussion}
The muti-state voter model with zealots can accurately estimate individual opinions in a large, fine-grained, heterogeneous online population. The separation between users with different ground-truth opinions is clear in the opinion space of the model, and the model correctly identifies the ground-truth for 92.5\% of the users. Discord probabilities act as very good proxies to distinguish user pairs with the same ground-truth opinion from others. \blue{These results are largely due to the direct connections with zealots, which act on their own as very good indicators of individual opinions. But the user-to-user influence mechanism encapsulated by the voter model allows for a greater accuracy, while extending the realm of the analysis to users who are not necessarily connected with zealots. Moreover, we have shown that networks derived from other forms of Twitter interactions (follow, mention) and unweighted, undirected counterparts of the retweet network do not yield as good results. This highlights the necessity to use specific types of interactions, and to not discard weights and directionality, for the voter model to retrieve individual opinions the best. These results advocate for the pursuit of more research dedicated to directed, weighted, heterogeneous networks with more than two opinions.} 

Our results also add to the existing body of evidence for the empirical validity of the voter model. Other datasets, and extensions such as the partisan voter model \cite{llabres2023partisan} or the nonlinear voter model \cite{ramirez2024ordering}, could be the ground of further testing. Because we analyzed the equilibrium state of the model, our findings pertain to the long-term behavior or the model. Future works should strive to integrate temporal aspects. Overall, we call for more research dedicated to the empirical evaluation of opinion dynamics models on the individual level in heterogeneous settings.

\begin{acknowledgments}
     I thank Fernando Diaz-Diaz for his careful reading and precious advices on this manuscript. I also thank Guillaume Cabanac and Oph\'elie Fraisier-Vannier for reuploading the \elysee\ dataset, Effrosyni Papanastasiou for sharing the follower network with me, and Luc\'ia S.\ Ramirez for insightful discussions on the voter model.
\end{acknowledgments}

\bibliographystyle{apsrev4-2}
\bibliography{biblio}

\end{document}



\title{Voter model can accurately predict individual opinions in online populations: Supplemental Material}

\author{Antoine Vendeville}
\email{antoine.vendeville@sciencespo.fr}
\affiliation{m\'edialab, Sciences Po, 75007 Paris, France}
\affiliation{Complex Systems Institute of Paris Île-de-France (ISC-PIF) CNRS, 75013 Paris, France}
\affiliation{Learning Planet Institute, Research Unit Learning Transitions, 75004 Paris, France}

\date{\today}
             
\maketitle

\section{Details of computations}
The values of $x_i^s$ and $\rho_{ij}$ are computed with iterative algorithms described in ref.~\cite{vendeville2023discord}. The argmax of $x_i$ and $z_i$ may lie in several of their coordinates at the same time ; when comparing these argmax with the ground-truth opinion, we consider that the ground-truth is correctly identified if one of the coordinates where the argmax is reached matches the ground-truth. The support vector classifier and the logistic regression are performed using the dedicated functions of Python's \texttt{sklearn} library. We use default parameters ; grid-searching for optimal values did not yield significantly better results. We use \texttt{class\_weight=balanced} for the logistic regression, because some parties have more supporters than others (cf.\ Fig~\ref{data_party_distrib}) and logistic regression is sensitive to class imbalance. Smoothed distributions are plotted using the \texttt{kdeplot} function of Python's \texttt{seaborn} library with default parameters.

\section{Statistics of networks}
\begin{figure}[!t]
     \centering
     \begin{subfigure}[b]{0.32\textwidth}
          \centering
          \includegraphics[width=\textwidth]{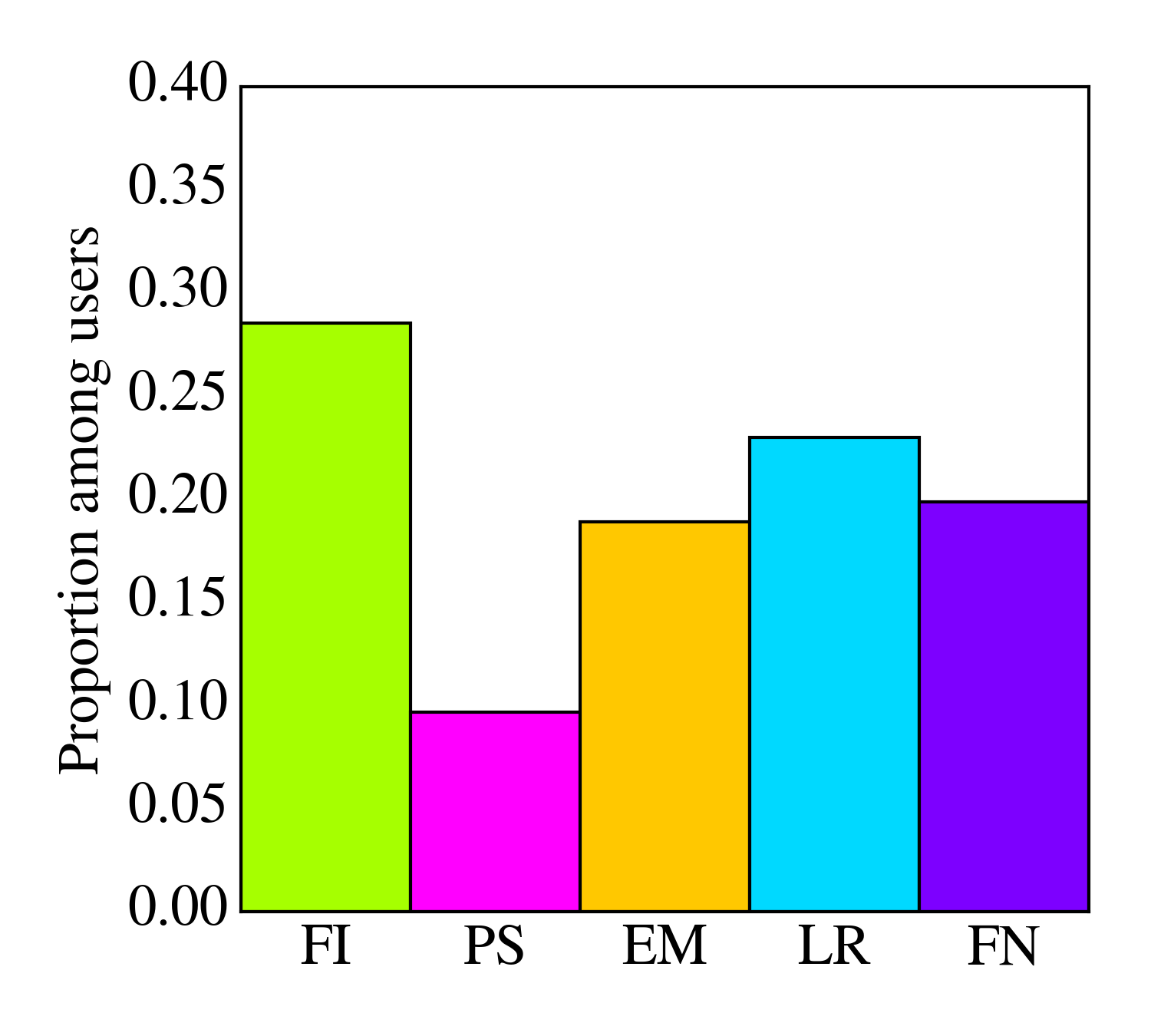}
     \end{subfigure}~
     \begin{subfigure}[b]{0.32\textwidth}
          \centering
          \includegraphics[width=\textwidth]{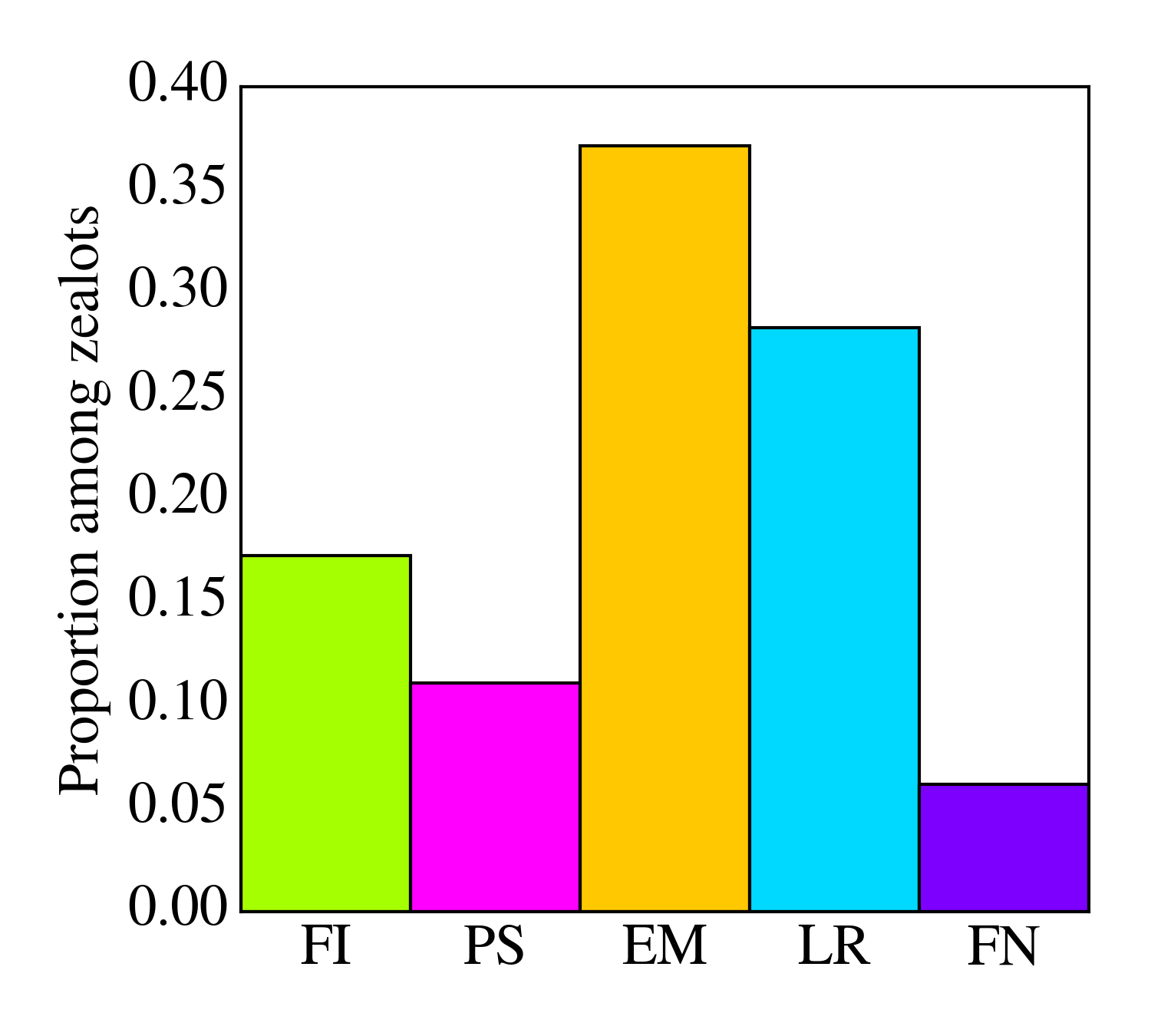}
     \end{subfigure}~
     \begin{subfigure}[b]{0.32\textwidth}
          \centering
          \includegraphics[width=\textwidth]{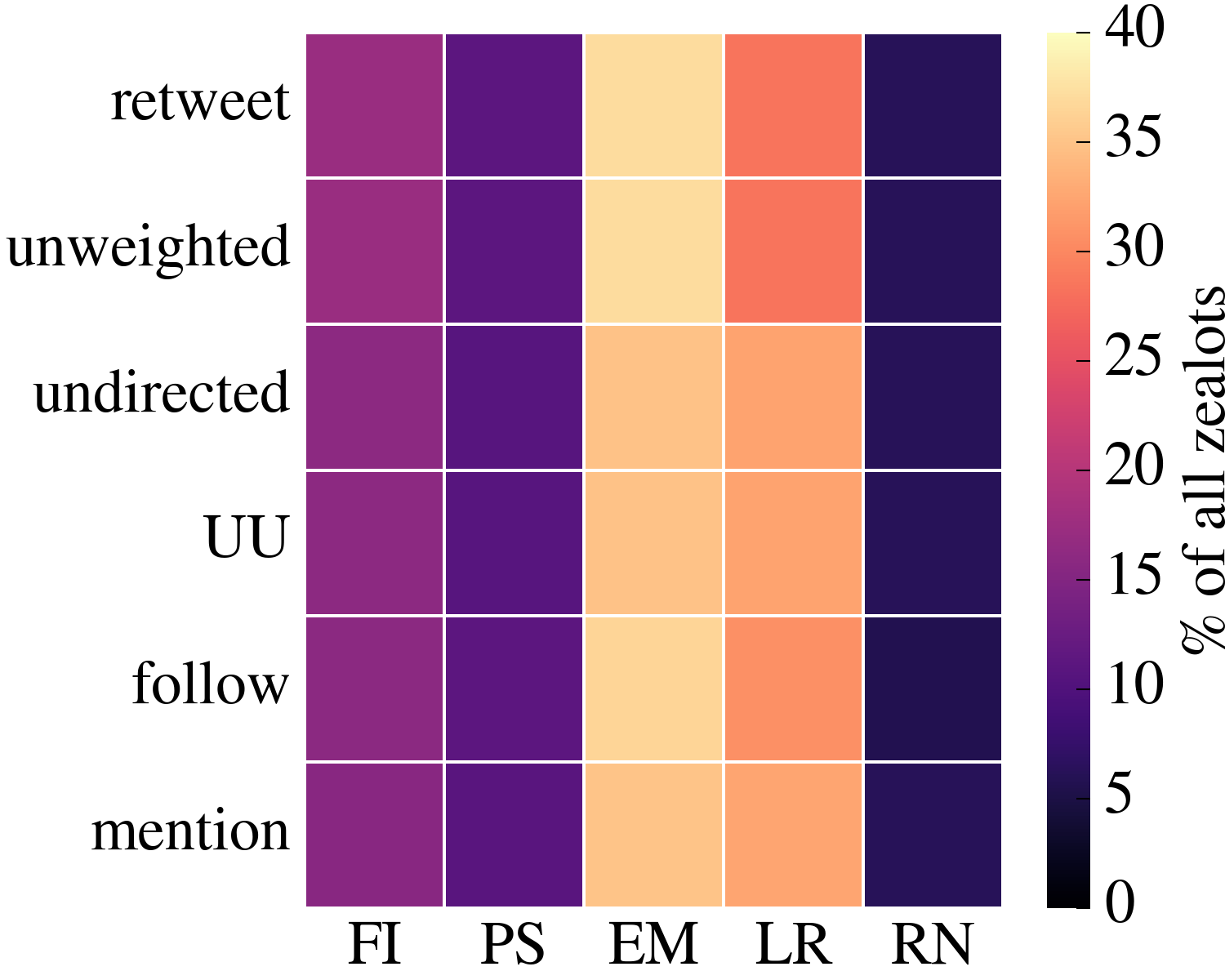}
          \vspace{-14pt}
     \end{subfigure}
     \caption{Distribution of ground-truth opinions in the retweet dataset, for users (left) and zealots (center). We also show the distribution of zealots across parties for each network (right). The rows are independent and each one sums to 100\%.}
     \label{data_party_distrib}
\end{figure}

The existence of a unique solution to Eq.~1 requires that every user can be reached by at least one zealot via a path in the network \cite[Thm.~2.1]{yildiz2013binary}. For each network considered (retweet, Unw., Und., UU, Follow, Mention) we remove all the users who cannot be reached by a zealot. Each network also exhibits a giant (weakly) connected component and several very small components of size one or two. Because our purpose is to study the effect of the user-to-user influence mechanism encapsulated by the voter model, we restrict ourselves to the giant component of each network. General statistics for the networks are provided in Table~\ref{tab:networks_stats}. The distribution of ground-truth labels for users and zealots in the retweet network is shown in Fig.~\ref{data_party_distrib}. The figures vary very little for the other networks. We also show the distribution of zealots across parties for each network. We see that all the networks exhibit highly similar distributions.

\begin{table}[t]
     \begin{tabular}{lcccccc}
          \hline
          & Retweet & Unw.\ & Und.\ & UU & Follow & Mention \\ \hline
          Number of users $N$ & 15,607 & 15,607 & 15,996 & 15,996 & 11,920 & 12,786 \\
          Number of zealots & 1,487 & 1,487 & 1,842 & 1,842 & 1,516 & 1,844 \\
          Percentage of zealots (\%) & 8.70 & 8.70 & 10.33 & 10.33 & 11.28 & 12.60 \\
          Size of the largest SCC & 10,618 & 10,618 & 15,996 & 15,996 & 10,385 & 12,618 \\ 
          Percentage of users in largest SCC (\%) & 68.03 & 68.03 & 100.00 & 100.00 & 87.12 & 98.69 \\
         \hline
     \end{tabular}
     \caption{Topological statistics of the different networks. SCC stands for Strongly Connected Component (of the user network).}
     \label{tab:networks_stats}
\end{table}

\begin{figure}[!t]
     \centering
     \includegraphics[width=\textwidth]{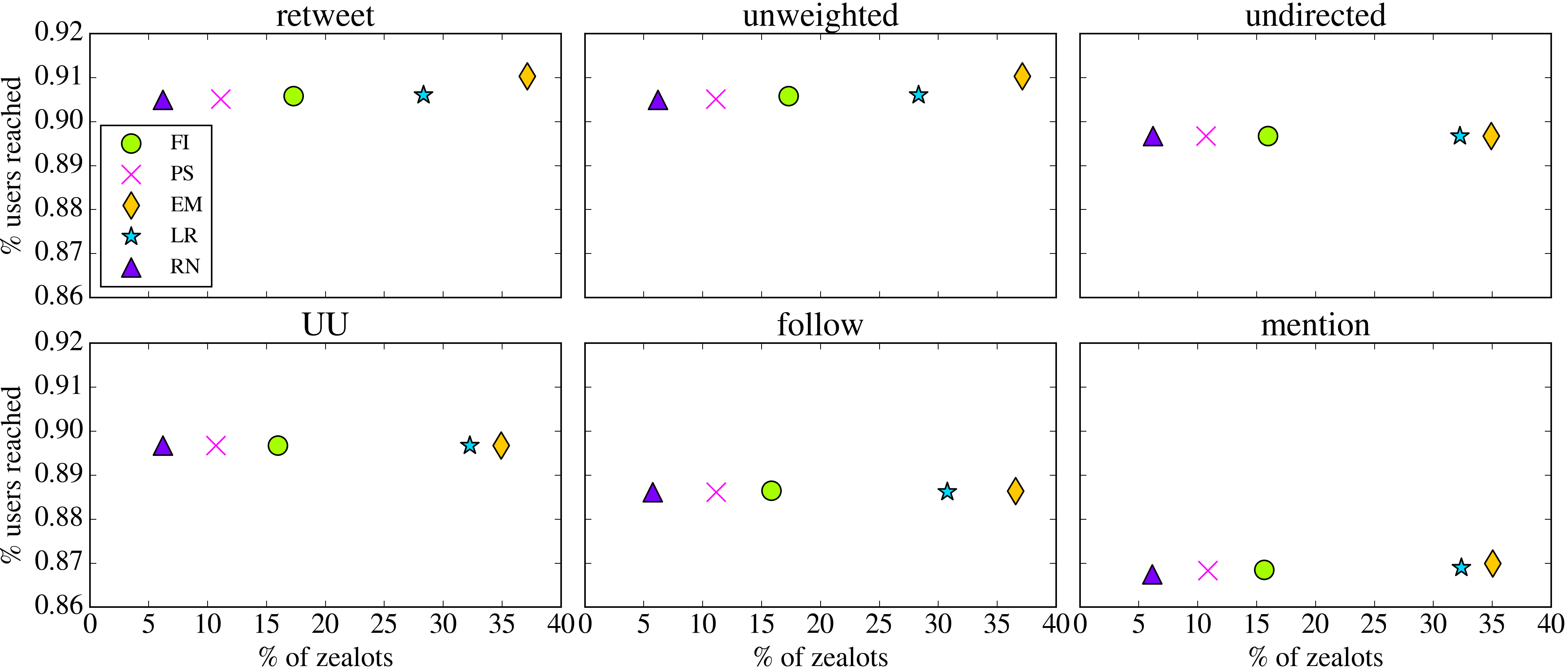}
     \caption{Comparison between the distribution of zealots across parties and their reach, defined for party $s$ as the number of users $i$ with $x_i^s>0$. We show results for each of the networks considered.}
     \label{reach_vs_percent_zealots}
\end{figure}

In Fig.~\ref{reach_vs_percent_zealots} we compare for each network, the distribution of zealots across parties with their reach. The reach of the zealots supporting party $s$ is defined as the number of users that they are able to reach, that is, the number of users $i$ with $x_i^s>0$. We do not see a strong difference in reach between parties in any network, however, we see that zealots have much less reach in the Follow and Mention networks. This is particularly interesting, given that these two networks boast the highest proportion of zealots among the whole population. Therefore, the evolution of opinions is more driven by interactions between users than by the influence of zealots, despite them being a larger share of the whole population. This could explain why the Follow and Mention networks exhibit the lowest performance in retrieving ground-truth opinions $y_i$ from the $x_i$ vectors.

\begin{figure}[!t]
     \centering
     \includegraphics[width=\textwidth]{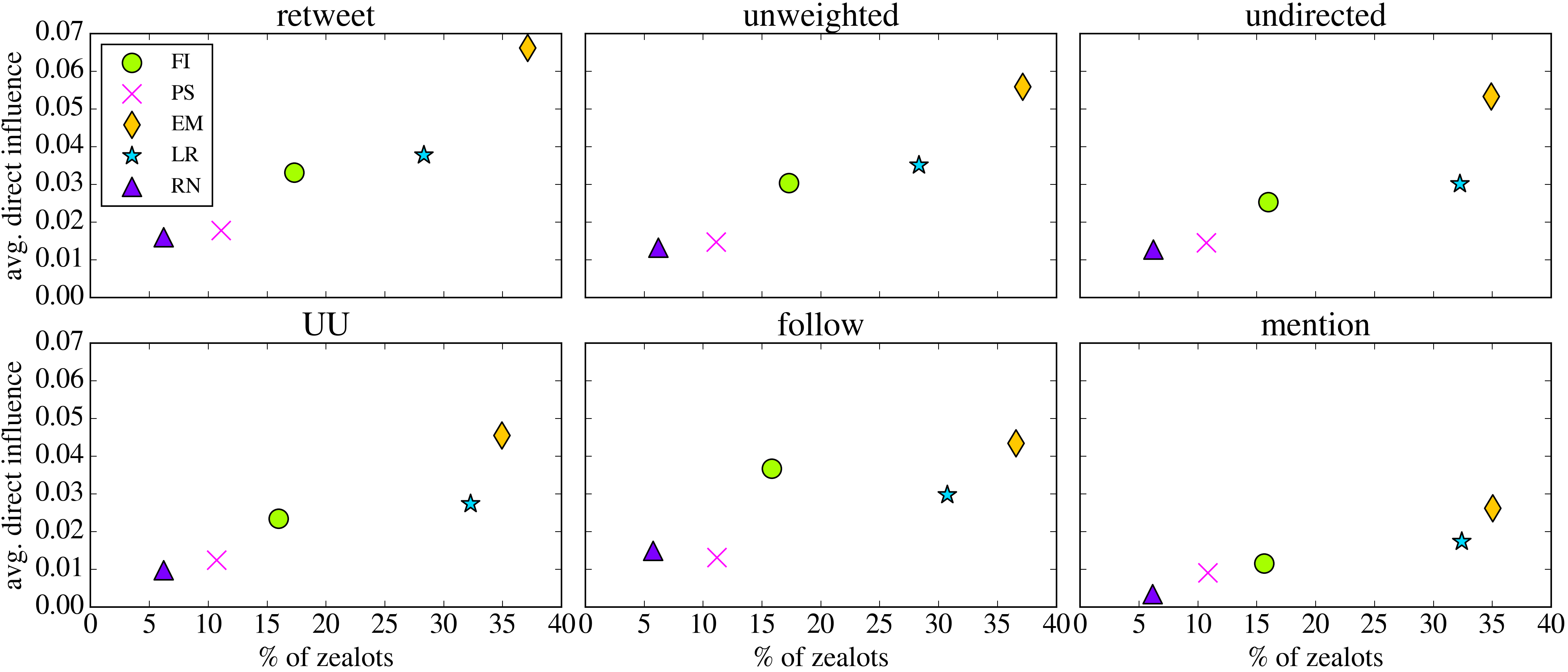}
     \caption{Comparison between the distribution of zealots across parties and their direct influence, defined for party $s$ as the average value of $z_i^s$. We show results for each of the networks considered.}
     \label{zis_vs_percent_zealots}
\end{figure}

We also look at the direct influence of zealots, defined for each party $s$ as the average value of $z_i^s$, in Fig.~\ref{zis_vs_percent_zealots}. This corresponds to a weighted out-degree centrality, shedding light on the importance of zealots within the networks. Again we do not see major differences between networks, except that zealots exhibit lower directed influences in the Mention network. Interestingly however, contrary to the reach, direct influences vary between parties. Parties with the largest number of zealots have more direct influence.

\section{Additional results and figures}

\subsection{Direct connections with zealots}

\begin{figure}[!t]
     \centering
     \includegraphics[width=\textwidth]{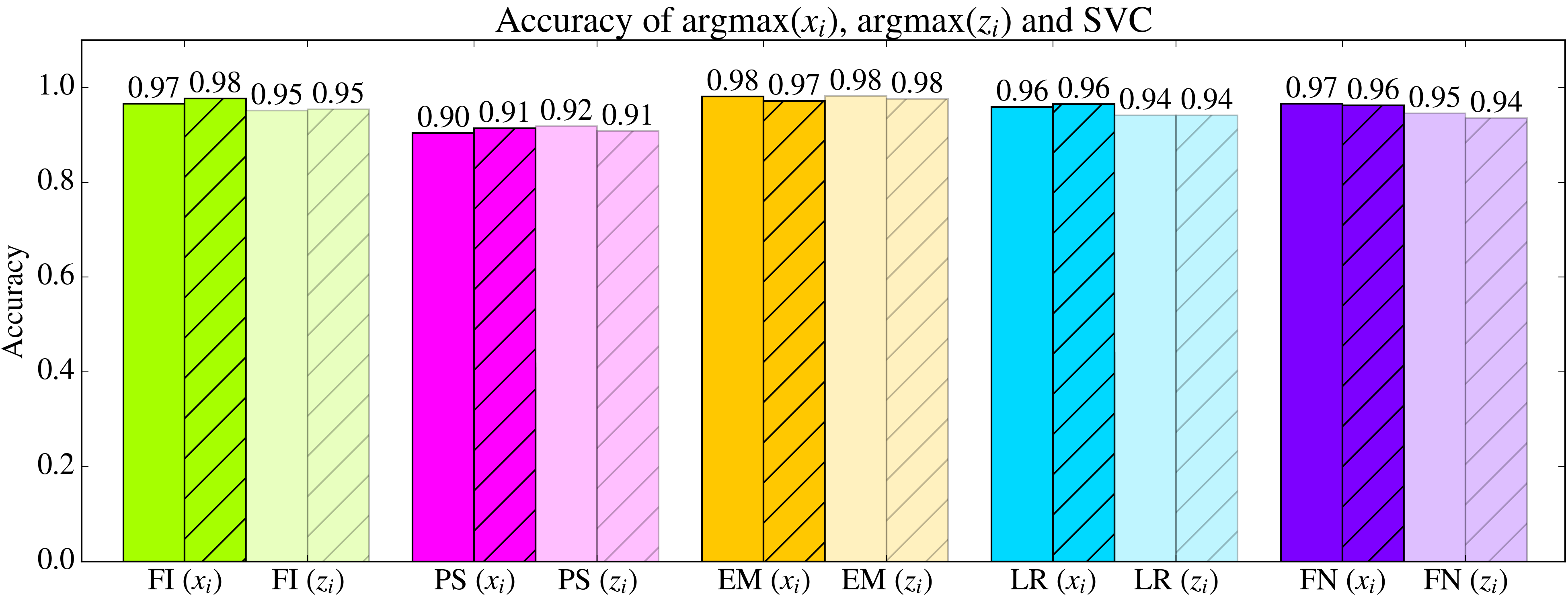}
     \caption{Party-wise accuracy of argmax$(x_i)$ (plain bars) and SVC (hatched bars) for the retrieval of ground-truth opinions $y_i$ from $x_i$ and $z_i$. Dark bars stand for $x_i$ and transparent bars for $z_i$.}
     \label{XvsZ}
\end{figure}

Figure~\ref{XvsZ} and Table~\ref{tab:x_vs_z} provide more detailed results about the comparison between the performance of $x_i$ and the performance of $z_i$. As stated in the main text, while users seem to be better separated along party lines in the opinion space by $z_i$ than by $x_i$ (last four lines of Table~\ref{tab:x_vs_z}), ground-truth opinions are more easily recognized by $x_i$ (first two lines of Table~\ref{tab:x_vs_z}).

\begin{table}[!t]
     \setlength{\tabcolsep}{12pt}
     \begin{tabular}{lcc}
          \hline
          & Voter model $(u_i=x_i)$ & Baseline $(u_i=z_i)$ \\ \hline
         Accuracy $\text{argmax} (u_i)$ & \textbf{96.2} & 95.1 \\
         Accuracy SVC & \textbf{96.4} & 94.8 \\
         $\langle u^s\rangle$ & 0.826 & \textbf{0.925} \\
         $\langle u^{-s}\rangle$ & 0.044 & \textbf{0.019} \\
         $\langle\Vert u_i-u_j\Vert\rangle_\text{within}$ & 0.219 & \textbf{0.190}  \\
         $\langle\Vert u_i-u_j\Vert\rangle_\text{cross}$ & 1.146 & \textbf{1.339}  \\
         \hline
     \end{tabular}
     \caption{Comparison between the equilibrium opinions in the voter model and a baseline that considers solely direct connections with zealots. The notation $u$ refers to $x$ for the voter model and to $z$ for the baseline. The network is restricted to users with at least one direct connection with zealots. Subscripts ``within'' and ``cross'' precise whether the average is taken over pairs of users supporting the same party or two different parties. Accuracies are given as percentages.}
     \label{tab:x_vs_z}
\end{table}

\subsection{Approximation of discord probabilities}

\begin{figure}[!t]
     \centering
     \includegraphics[width=\textwidth]{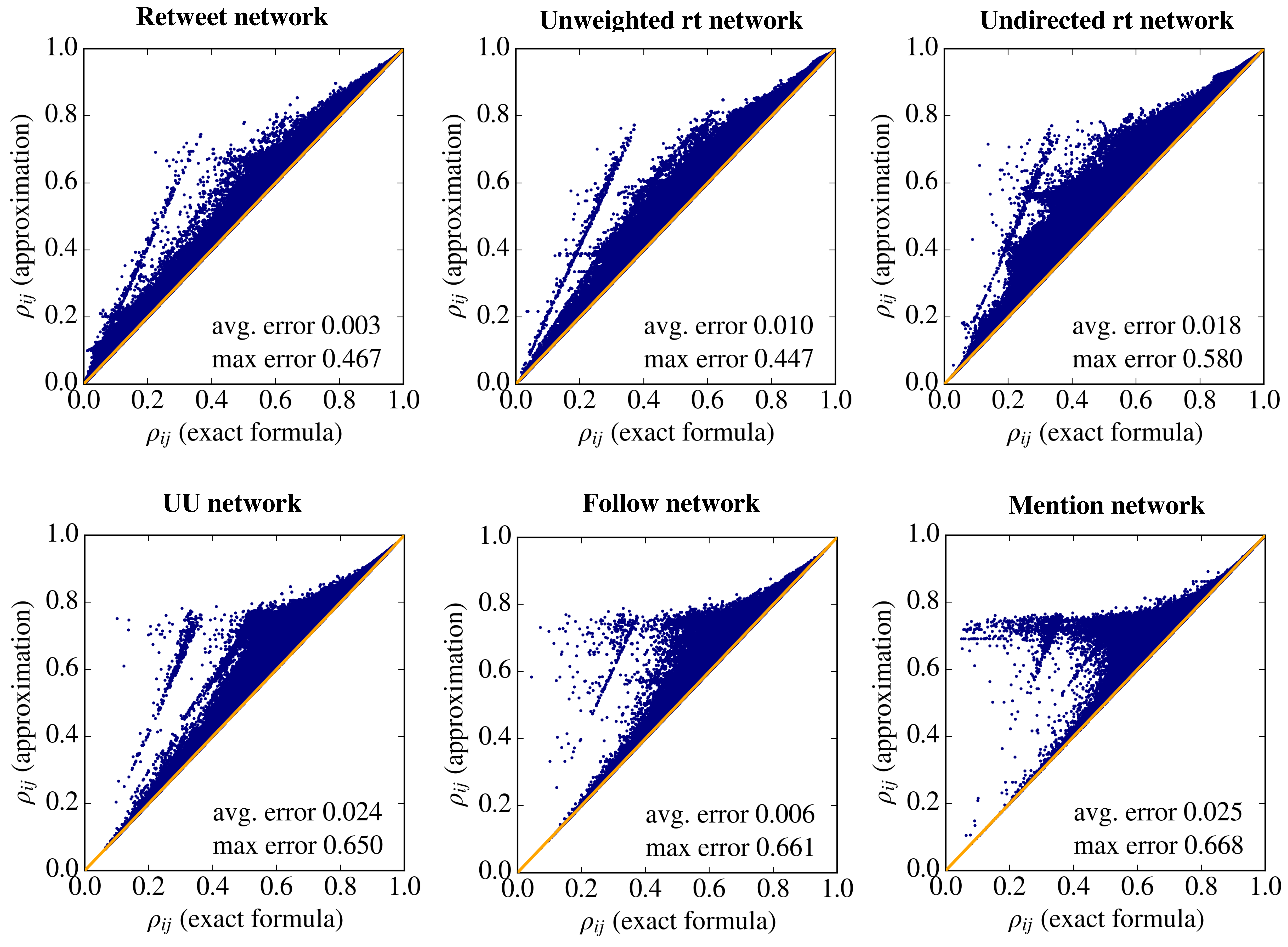}
     \caption{Comparison between $\rho_{ij}$ values computed using the exact formula (Eq.~5 in the main text, x-axis) and the approximation proposed in \cite[Eq.~10]{vendeville2023discord} (y-axis) for each network. The diagonal orange lines indicate the identity.}
     \label{dep_indep}
\end{figure}

We proposed in ref.~\cite[Eq.~10]{vendeville2023discord} a formula to approximate quickly and efficiently the discord probabilities $\rho_{ij}$ on the sole base of $x_i$ and $x_j$. The approximation becomes exact under certain conditions of independence between the opinions of $i$ and $j$. To assess the approximation error, we compare in Fig.~\ref{dep_indep} its results with those obtained from the exact formula (Eq.~5 in the main text). We disregard user pairs with independent opinions and for which the approximation is exact. We find that the approximation tends to over-estimate, and rarely under-estimate, the discord probabilities. This is not too surprising, given that factors of dependencies between opinions tend to bring those toward the same values. For example, if two users are connected by heavily weighted edges, their opinions will naturally be much more than could be assumed from $x_i$ and $x_j$ taken separately. The average error lies between 0.003 (Retweet network) and 0.025 (Mention network), with maximum errors ranging from  0.447 (unweighted retweet network) to 0.668 (Mention network). Therefore, while the approximation fares well for most users, if one is interested in the discord between two users which highly dependent opinions, it becomes crucial to use the exact formula.

\subsection{Results for additional networks}

\begin{figure}[!t]
     \centering
     \includegraphics[width=.9\textwidth]{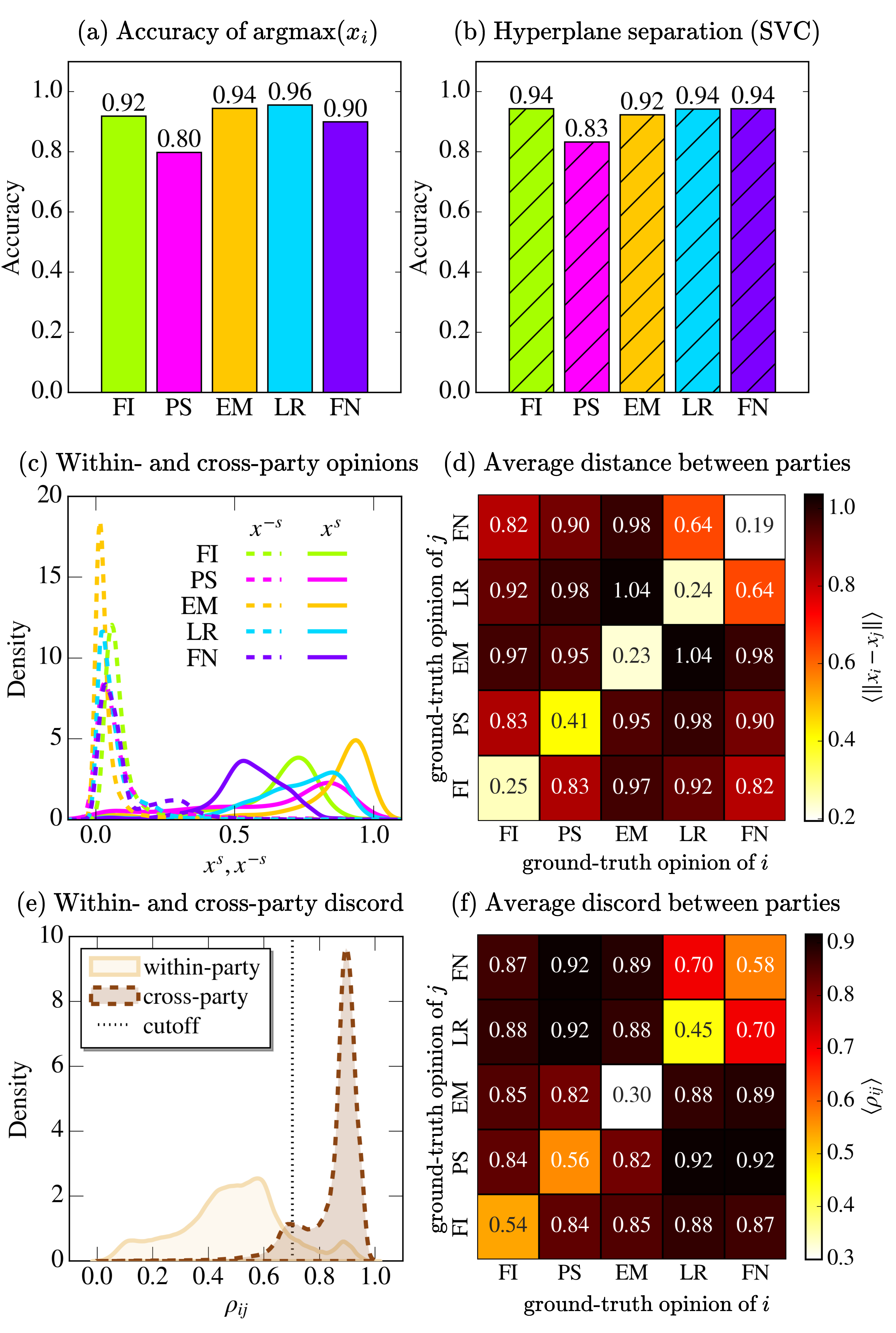}
     \caption{Correspondence between theoretical opinion distribution $x_i$ and ground-truth opinions $y_i$ for the unweighted retweet network.}
     \label{assemble_unweighted}
\end{figure}

\begin{figure}[!t]
     \centering
     \includegraphics[width=.9\textwidth]{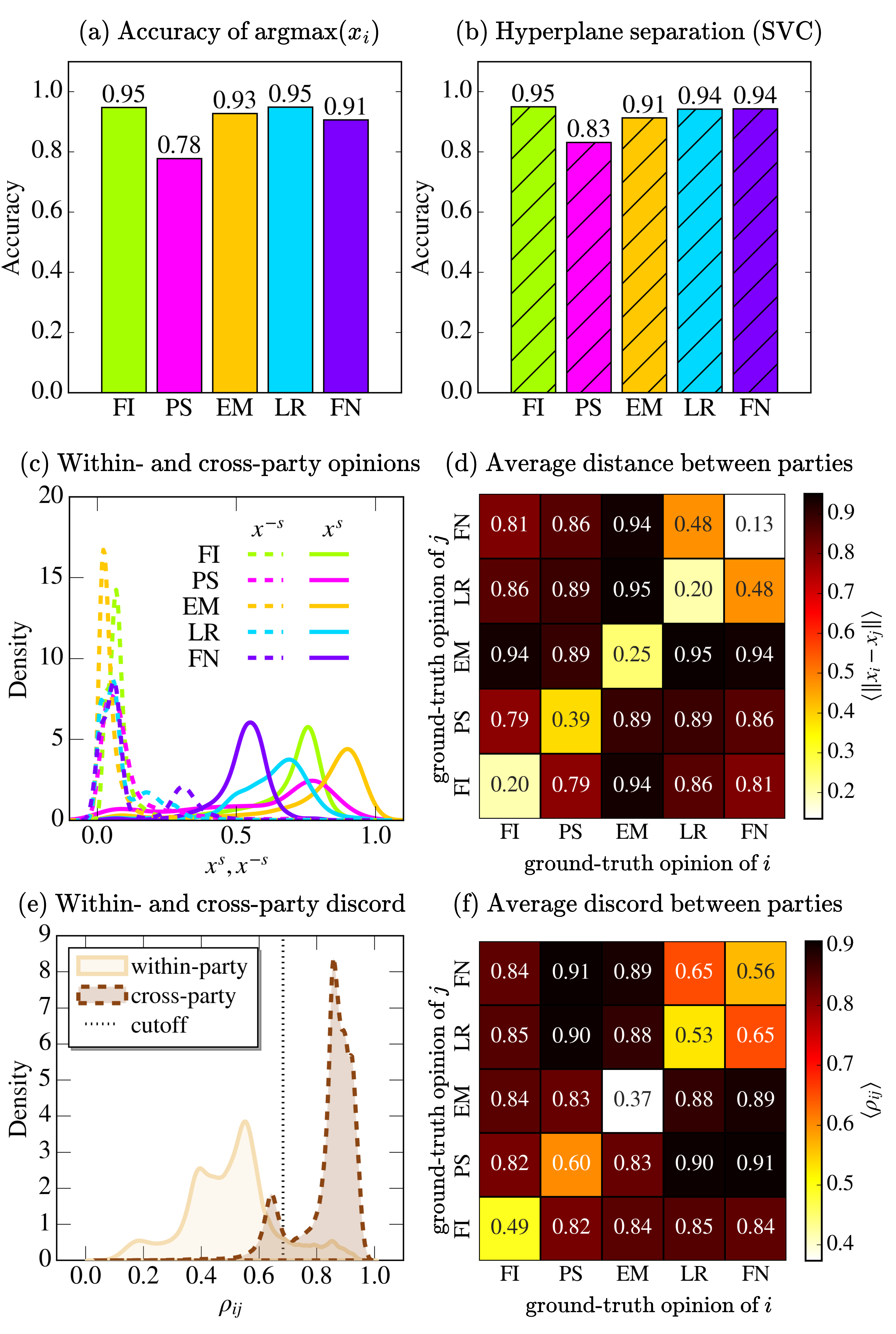}
     \caption{Correspondence between theoretical opinion distribution $x_i$ and ground-truth opinions $y_i$ for the undirected retweet network.}
     \label{assemble_undirected}
\end{figure}

\begin{figure}[!t]
     \centering
     \includegraphics[width=.9\textwidth]{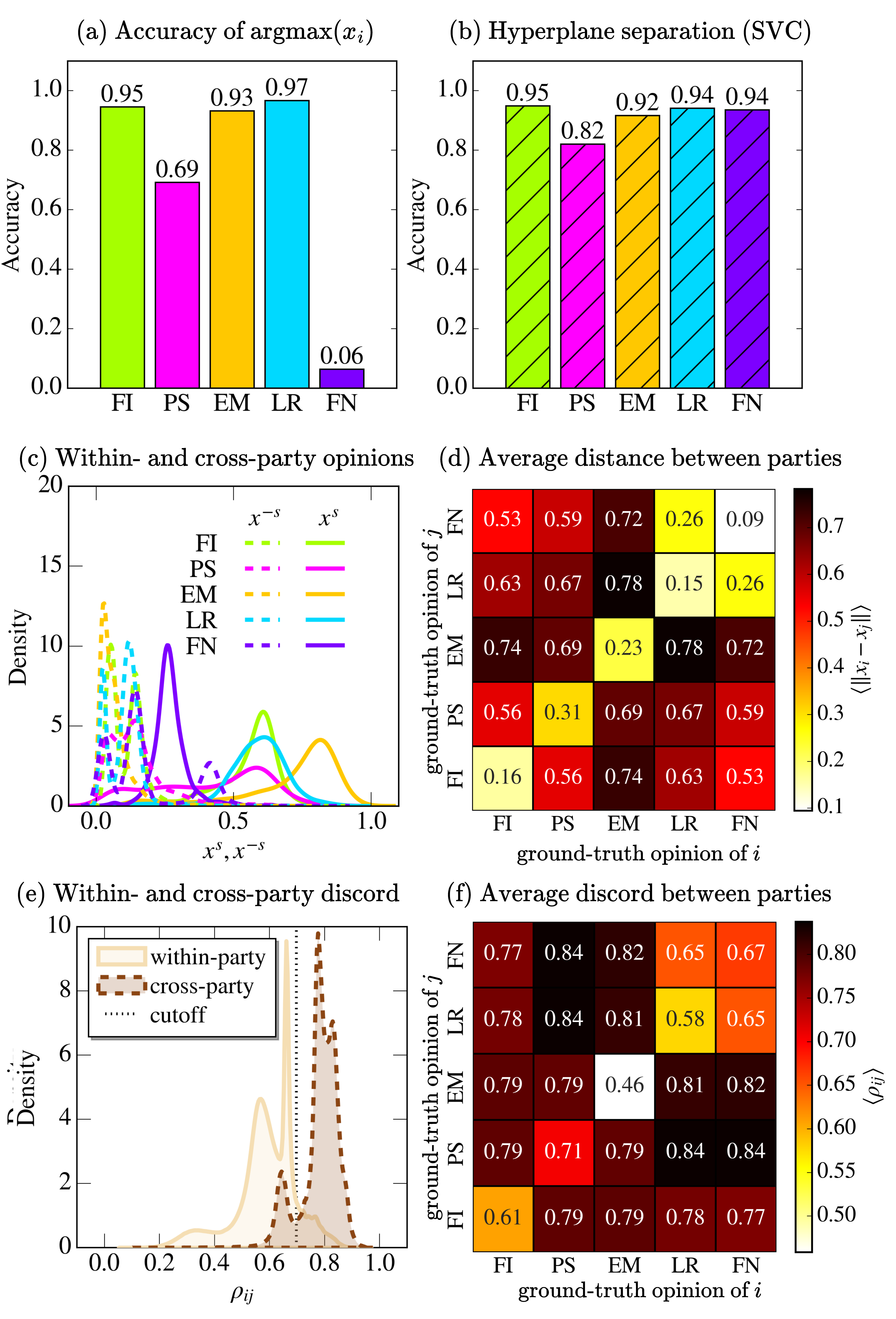}
     \caption{Correspondence between theoretical opinion distribution $x_i$ and ground-truth opinions $y_i$ for the undirected and unweighted (UU) retweet network.}
     \label{assemble_UU}
\end{figure}

\begin{figure}[!t]
     \centering
     \includegraphics[width=.9\textwidth]{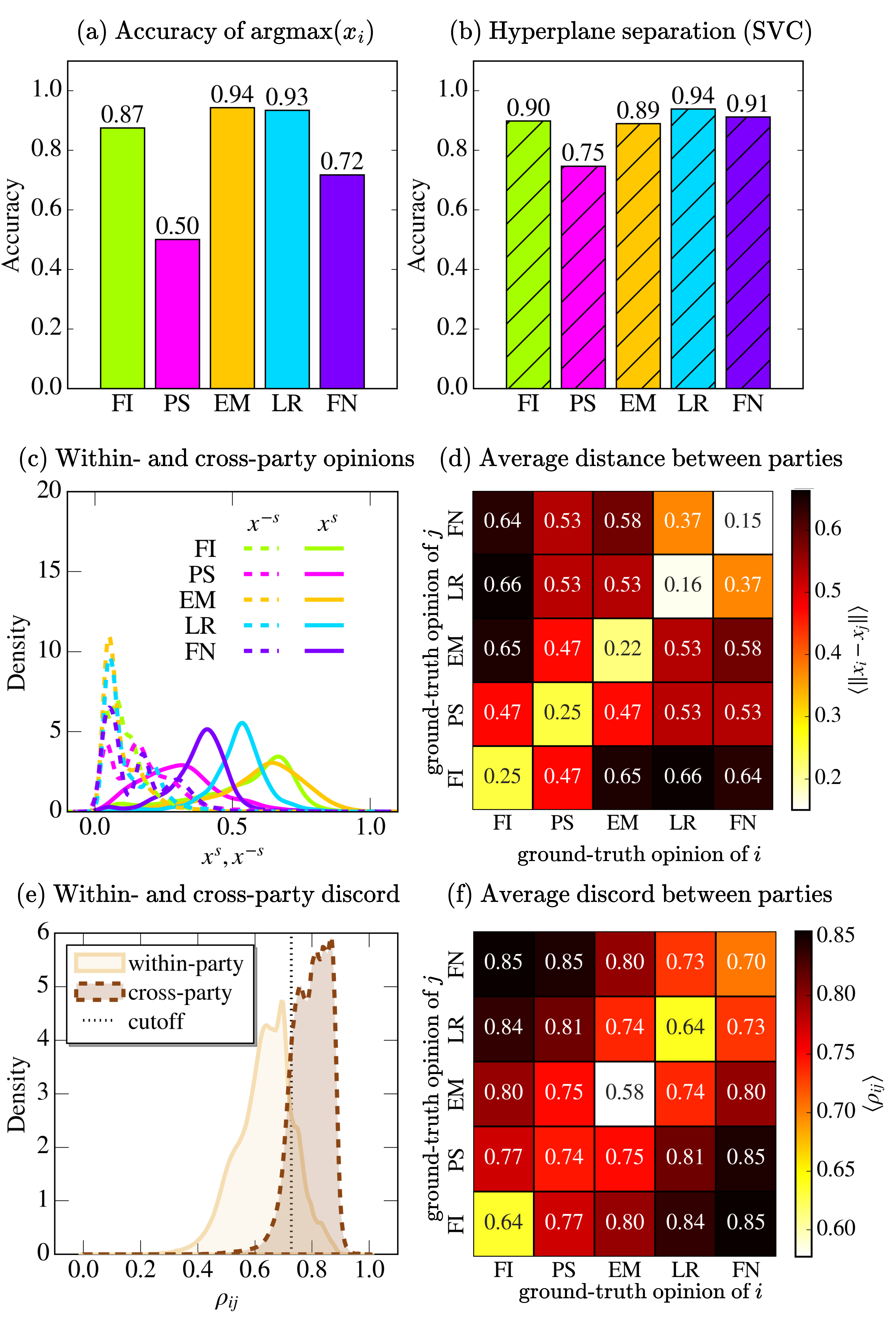}
     \caption{Correspondence between theoretical opinion distribution $x_i$ and ground-truth opinions $y_i$ for the Follow network.}
     \label{assemble_follow}
\end{figure}

\begin{figure}[!t]
     \centering
     \includegraphics[width=.9\textwidth]{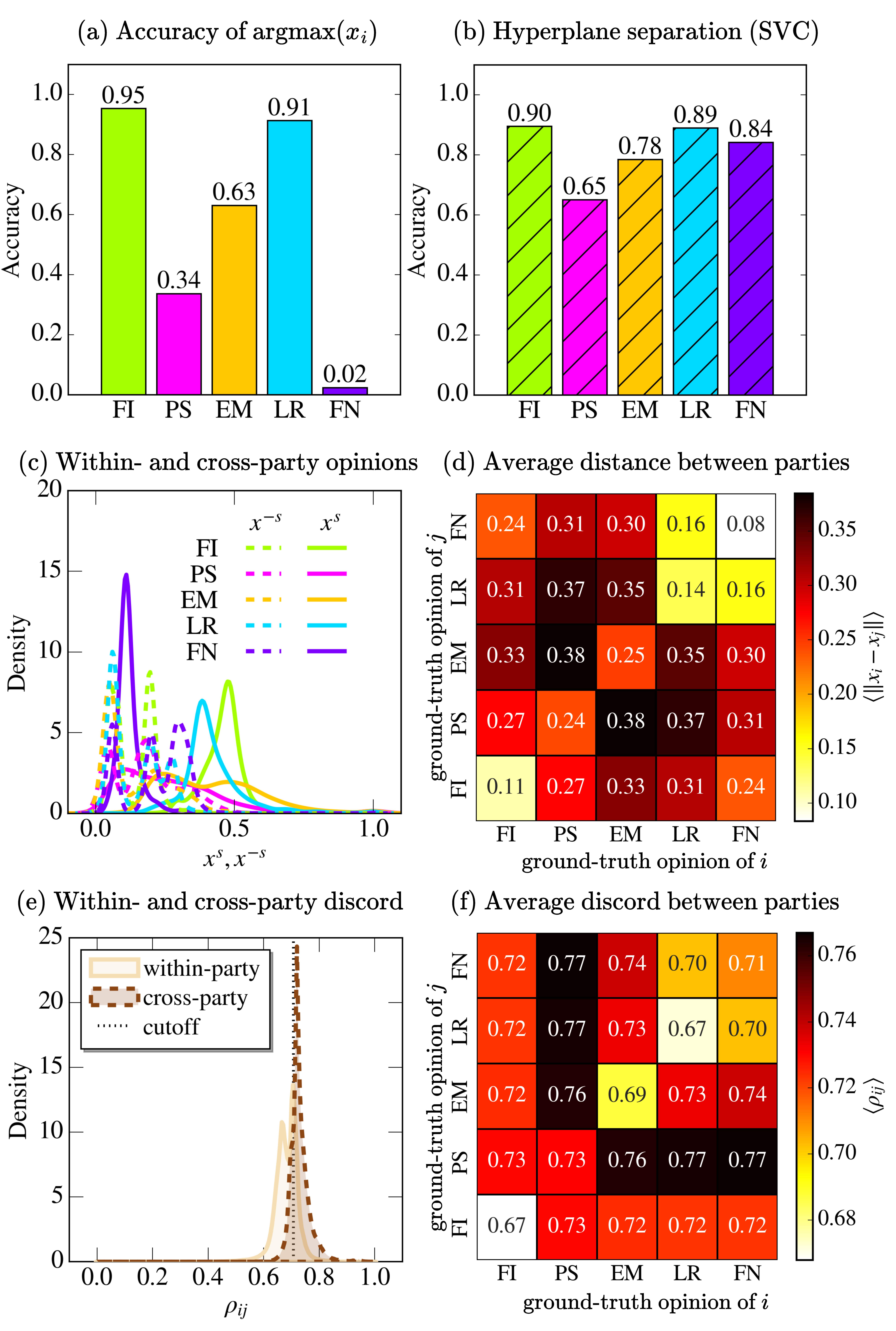}
     \caption{Correspondence between theoretical opinion distribution $x_i$ and ground-truth opinions $y_i$ for the Mention network.}
     \label{assemble_mention}
\end{figure}

We show the equivalent of Fig.~1 of the main text for the other networks: Unw.\ (Figs.~\ref{assemble_unweighted}), Und.\ (Fig.~\ref{assemble_undirected}), UU (Fig.~\ref{assemble_UU}), Follow (Fig.~\ref{assemble_follow}), Mention (Fig.~\ref{assemble_mention}). 

\bibliographystyle{apsrev4-2}
\bibliography{biblio}